\documentclass[noshowpacs,amsmath,
 twocolumn,
 superscriptaddress,
 8pt,
 aps,prb]{revtex4-2}
\usepackage{setspace}
\usepackage{amsmath}
\usepackage{multirow}
\usepackage{float}
\usepackage{graphicx}

\usepackage{verbatim}
\usepackage{amsfonts}
\usepackage{amssymb}
\usepackage{epstopdf}
\usepackage{dcolumn}
\usepackage{xcolor}
\usepackage{verbatim}
\usepackage{hyperref}
\usepackage{siunitx}
\usepackage[version=4]{mhchem}
\usepackage{textcomp}
\setcitestyle{super} 
\DeclareGraphicsExtensions{.pdf,.eps,.png,.jpg,.mps}

\begin{document}

\title{Communication-ready high-power soliton microcombs in highly-dispersive Fabry-Pérot-microresonators}

\author{Yinke Cheng$^{1,\dagger}$, Zhenyu Xie$^{1,\dagger}$, Yuanlei Wang$^{1,\dagger}$, Binbin Nie$^{1}$, Xing Jin$^{1}$, Haoyang Luo$^{1}$, Junqi Wang$^{1}$, Zixuan Zhou$^{5}$, Qihuang Gong$^{1,2,3,4}$, Lin Chang$^{5}$, Yaowen Hu$^{1,2,3}$, and Qi-Fan Yang$^{1,2,3,4,*}$\\
$^1$State Key Laboratory for Artificial Microstructure and Mesoscopic Physics and Frontiers Science Center for Nano-optoelectronics, School of Physics, Peking University, Beijing 100871, China\\
$^2$Peking University Yangtze Delta Institute of Optoelectronics, Nantong, Jiangsu 226010, China\\
$^3$Collaborative Innovation Center of Extreme Optics, Shanxi University, Taiyuan 030006, China\\
$^4$Hefei National Laboratory, Hefei 230088, China\\
$^5$State Key Laboratory of Advanced Optical Communications System and Networks, School of Electronics, Peking University, Beijing, 100871, China\\
$^{\dagger}$These authors contributed equally to this work.\\
$^{*}$Corresponding author: leonardoyoung@pku.edu.cn}
% use {asbstract*} to suppress the copyright line. Copyright information will be added in production

\begin{abstract}
Microcombs generated in optical microresonators are widely regarded as promising light sources for next-generation communication systems, but the optical power available per comb line has so far fallen short of practical requirements. Here we introduce an integrated Fabry–Pérot microresonator platform that overcomes fundamental dispersion-engineering constraints and enables bright soliton microcombs with unprecedented power per line. The resonator is defined by chirped Bragg gratings that provide exceptionally large anomalous group-velocity dispersion, allowing more than ten comb lines to reach the milliwatt level. These combs can be used directly in coherent communication systems without additional amplification, achieving an aggregate data rate of 2~Tb/s. Once integrated, our high-power soliton microcombs could be instantly ready for communications as well as a broad range of practical comb-based applications.
\end{abstract}

\maketitle

\section{Introduction}
Escalating traffic from cloud and artificial-intelligence workloads is driving a major upgrade of optical communication systems \cite{agrell2016roadmap}. On the transmitter side, co-packaged optics (CPO) are being introduced to replace conventional pluggable modules, thereby improving both energy efficiency and symbol rate \cite{Margalit2021Perspective}. In parallel, a transformation is underway in the light sources that carry the information \cite{chang2022integrated}. These sources must provide many wavelength channels and support advanced modulation formats in order to maximize the throughput per fiber. However, implementing large arrays of distributed-feedback (DFB) lasers incurs severe penalties in footprint, power consumption, and thermal management. Kerr frequency combs generated in high-$Q$ microresonators offer a scalable alternative \cite{Kippenberg2018}: a single continuous-wave (CW) pump laser can generate tens to hundreds of equally spaced lines that are naturally compatible with wavelength-division-multiplexing grids. Such Kerr microcombs have been tested extensively in a wide range of communication scenarios, including long-haul coherent links \cite{marin2017microresonator,corcoran2020ultra,jorgensen2022petabit,Geng2022Coherent} and short-reach direct-detection interconnects \cite{shu2022microcomb,Yang2022Multidimensional,rizzo2023massively}.

Moving from laboratory demonstrations to deployed systems imposes stringent requirements on the power available per comb line. Typically, to drive a CPO engine, the optical power per wavelength for the external laser source is a few milliwatts \cite{Buscaino2021ExternalIntegrated}. Dark-pulse microcombs operating in the normal group-velocity-dispersion (GVD) regime are attractive because they can provide relatively high line powers, but their spectral profiles are difficult to control and remain challenging to predict \cite{xue2015mode,kim2019turn,helgason2021dissipative,jin2021hertz,zang2025laser}. By contrast, bright-soliton microcombs in anomalous GVD microresonators exhibit well-defined $\mathrm{sech}^2$ spectral envelopes that are favorable for system design, yet the power per comb line is generally low \cite{herr2014temporal,yi2015soliton,brasch2016photonic,wang2018robust,gong2018high,he2019self,wang2024lithium,nie2025soliton,song2025stable}. Conventional bright-soliton microcombs typically provide comb lines below $-10$~dBm, necessitating additional amplification in communication experiments \cite{marin2017microresonator}. These power constraints complicate communication architectures and hinder comb-based implementations in other power-hungry applications, such as ranging (Fig. 1(a)).

\begin{figure*}[t]
\centering\includegraphics[width=\linewidth]{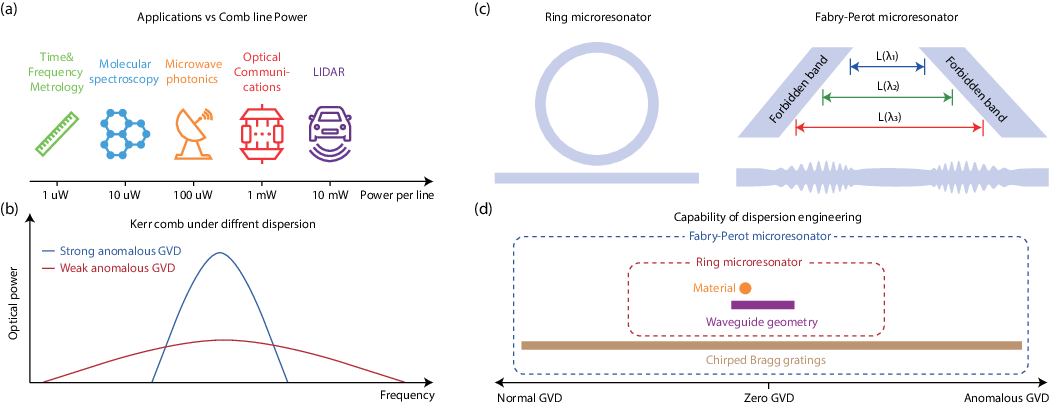}
\caption{{\bf Applications and implementation of high-power soliton microcombs.} 
(a) Representative microcomb-enabled applications and their required power per comb line. 
(b) Conceptual spectra of microcombs under different group-velocity-dispersion (GVD) conditions. 
(c) Schematic comparison between a ring microresonator (left) and a chirped Bragg-grating Fabry-Pérot (FP) microresonator (right). 
(d) Illustration of the dispersion-engineering degrees of freedom and accessible range afforded by FP versus ring microresonator geometries.}
\end{figure*}

Like solitons in optical fibers, the power of soliton microcombs is governed by a balance between dispersion and Kerr nonlinearity. A simple scaling relation for the power in a central comb line is
\begin{equation}
P_c \propto \frac{D_2}{Q},
\end{equation}
where $D_2$ characterizes the GVD of the resonator and $Q$ is the loaded quality factor \cite{yi2015soliton}. Decreasing $Q$ increases $P_c$, but also raises the threshold for parametric oscillation as $1/Q^2$ \cite{kippenberg2004kerr}. A more favorable route is therefore to increase $D_2$, which boosts the comb-line power while narrowing the optical bandwidth (Fig.1(b)). In practice, however, standard ring microresonators offer limited freedom for GVD engineering because material dispersion and geometric confinement restrict the accessible parameter space. We address this limitation by introducing Fabry-Pérot (FP) microresonators as an alternative platform for high-power Kerr microcombs (Fig.1(c)). In these devices, dispersion is engineered through photonic-bandgap physics rather than solely through waveguide geometries \cite{wildi2023dissipative,nardi2024integrated}. Reflection from integrated Bragg gratings defines the resonator, and by chirping the Bragg gratings, the resonator's round-trip length becomes a designed function of wavelength, providing an additional and highly flexible degree of freedom. This control dramatically enlarges the accessible GVD parameter space (Fig.1(d)) and, as we show below, enables soliton microcombs with exceptionally high comb-line powers.

\section{Microresonator design}

\begin{figure*}[t]
\centering\includegraphics[width=\linewidth]{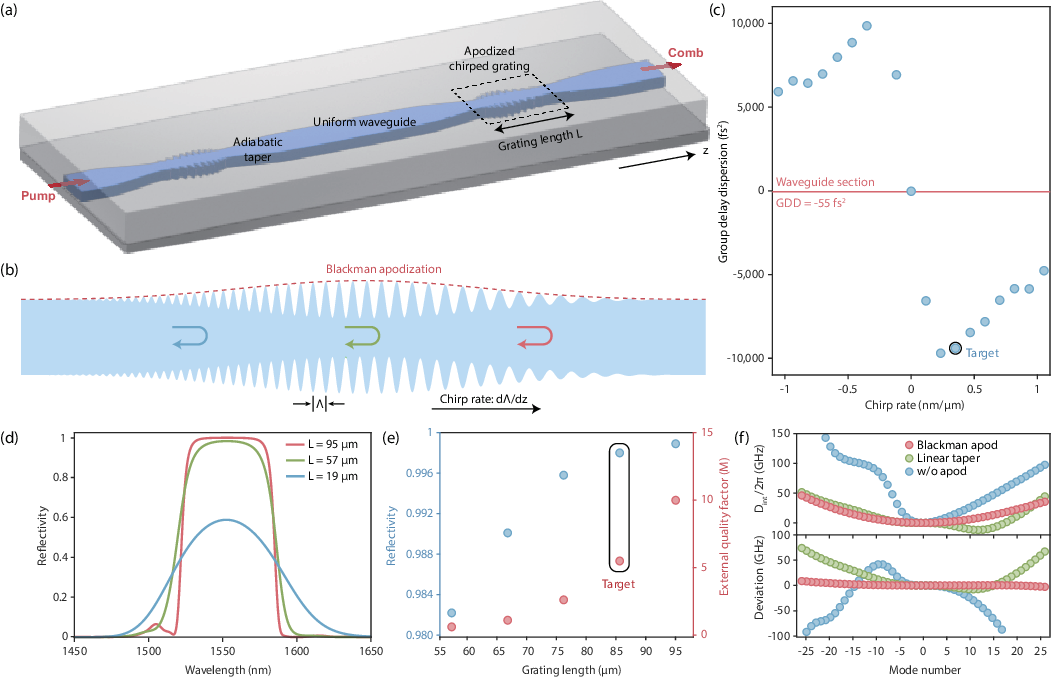}
\caption{{\bf Design of an integrated chirped Bragg-grating Fabry--Pérot microresonator.} 
(a) Schematic of the integrated FP microresonator formed by two chirped Bragg gratings (CBGs) and a uniform waveguide section. (b) Enlarged view of a CBG, whose modulation depth is apodized using a Blackman function. (c) Simulated dependence of the GVD on the chirp rate. (d) Reflectivity of the Bragg grating as a function of grating length. (e) Simulated coupling $Q$ factor versus grating length. (f) Integrated dispersion and its deviation from a purely quadratic profile for different apodization schemes.}
\end{figure*}

Although the FP concept is generic and can be applied to a broad range of Kerr-comb platforms, here we implement it on a high-confinement Si$_3$N$_4$ platform (777-nm-thick Si$_3$N$_4$ core with silica cladding), chosen for its ultralow propagation loss and wide transparency window \cite{ji2017ultra,Liu2021HighYieldSiN,jin2021hertz,Xiang2022SiNtrends,Buzaverov2024SiNvisibleMidIR}. The device layout (Fig.2(a)) consists of an FP microresonator formed by two chirped Bragg gratings separated by a uniform waveguide section. The uniform waveguide has a length of 500~$\mu$m and a width of 2~$\mu$m. Sidewall-modulated integrated Bragg gratings serve as mirrors, providing strong and controllable index contrast and low excess loss \cite{tan2009cladding,hung2015narrowband,huang2016ultra,li2020ultra}. To minimize transition loss between the gratings and the uniform waveguide, a 100-$\mu$m-long adiabatic taper is inserted. The free spectral range (FSR) of the microresonator is approximately 100~GHz.

The CBGs employ sinusoidal sidewall modulation with apodized amplitude profiles (Fig.2(b)). The average waveguide width in the grating region is 0.6~$\mu$m to ensure single-mode operation and suppress higher-order transverse modes. The maximum modulation depth is 0.3~$\mu$m, providing sufficient reflectivity per unit length to reduce the device footprint without incurring excessive scattering loss. The central period $\Lambda$ fixes the center wavelength of the photonic bandgap; we set $\Lambda = 0.466~\mu$m so that the reflection band is centered near 1550~nm in the telecommunications C-band.

The dispersion of a longitudinal mode family in a microresonator is conveniently expressed in terms of the integrated dispersion,
\begin{equation}
D_\mathrm{int}(\mu) = \omega_\mu - \mu D_1 = \frac{\mu^2 D_2}{2} + \mathcal{O}(\mu^3),
\end{equation}
where $\omega_\mu$ is the resonance frequency of the mode with index $\mu$ (with $\mu=0$ at the pump), and $D_1/2\pi$ is the FSR. For high-power soliton microcomb generation, $D_2$ must be strongly anomalous (positive in value) while higher-order dispersion remains small. In the FP microresonator, $D_2$ receives contributions from the material and geometric dispersion of the uniform waveguide as well as from the CBGs, which introduce a wavelength-dependent group delay \cite{senior2009optical,komukai2002design}. We evaluate the group-delay dispersion (GDD) of each section, which determines $D_2$ via
\begin{equation}
D_2 = -\frac{D_1}{2\pi}\,\mathrm{GDD}.
\end{equation}
For the CBGs, the GDD at wavelength $\lambda$ is approximately related to the chirp rate $\mathrm{d}\Lambda/\mathrm{d}z$ by
\begin{equation}
\mathrm{GDD} \approx -\frac{\lambda^2}{2\pi c^2}\frac{1}{\mathrm{d}\Lambda/\mathrm{d}z},
\end{equation}
so that the magnitude of the GDD scales inversely with the chirp rate. At smaller chirp rates, however, variations in local reflectivity and penetration depth become important, and higher-order dispersion---particularly third-order dispersion---dominates near the band center. As the chirp tends to zero, the second-order dispersion vanishes at the reflection-band center and the GDD exhibits a non-monotonic dependence on chirp, first increasing and then decreasing. These behaviors are confirmed by numerical simulations (see Supplementary Materials). Based on this analysis, we choose a chirp rate of 0.35~nm/$\mu$m, corresponding to the largest negative GDD (Fig.2(c)). The resulting GDD of the CBG section is $-8997$~fs$^2$, more than 160 times larger in magnitude than that of the uniform waveguide section (55~fs$^2$).

The total grating length is optimized to balance reflection bandwidth and peak reflectivity. For the chosen chirp rate, we simulate the wavelength-dependent reflectivity as a function of grating length (Section 1 in Supplement). Increasing the grating length enhances the peak reflectivity and slightly modifies the usable bandwidth (Fig.2(d)). The reflectivity sets the coupling between the external bus waveguide and the FP microresonator, and therefore the external $Q$ factor. Given an intrinsic $Q$ of approximately $5\times10^6$ inferred from the waveguide propagation loss, we target a comparable external $Q$. Detailed simulations show that gratings with an effective length of about 85~$\mu$m yield an external $Q$ of $\sim 5\times10^6$, and this length is adopted (Fig.2(e)).

Apodization of the grating amplitude is essential for achieving a spectrally uniform reflection profile and suppressing higher-order dispersion. Abrupt changes in the refractive-index modulation at the grating edges act as local FP etalons, generating spectral ripples and strong higher-order dispersion \cite{cheng2020spectral}. We compare CBGs with Blackman apodization \cite{tan2008chip,kashyap2009fiber}, linear-taper apodization, and no apodization in terms of their impact on the integrated dispersion (Fig.2(f)). Blackman apodization yields integrated-dispersion curves that closely follow a parabola, indicating strongly suppressed higher-order dispersion and an almost ideal quadratic mode family.

\medskip 
\section{ Device characterization}

\begin{figure*}[t]
\centering\includegraphics[width=\linewidth]{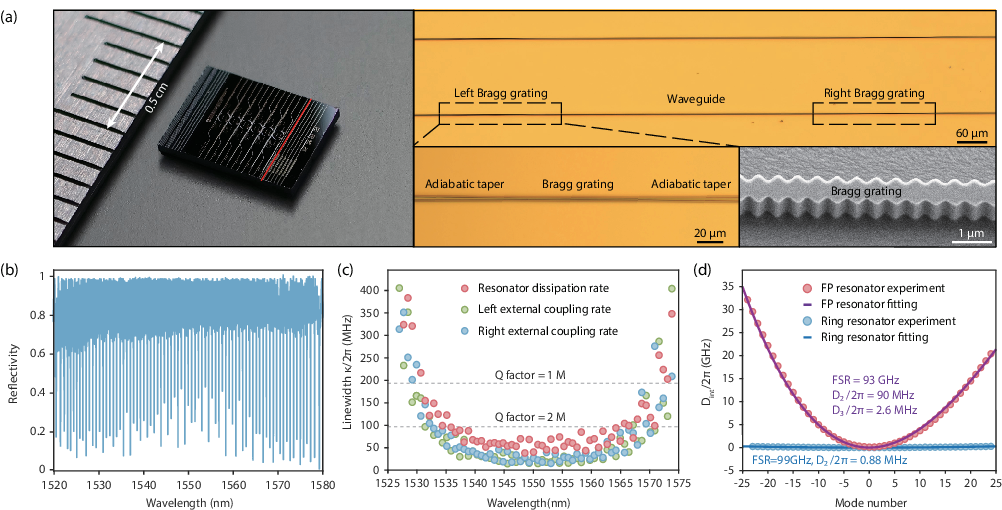}
\caption{{\bf Characterization of the integrated FP microresonator.} 
(a) Optical micrograph of the chip, zoomed-in view of the integrated FP microresonator, and scanning electron micrograph of the chirped Bragg grating. (b) Normalized reflection spectrum of the microresonator. (c) Intrinsic dissipation and external coupling rates of the microresonator. (d) Measured integrated dispersion of the FP microresonator compared to a ring resonator with the same waveguide cross-section.}
\end{figure*}

The devices are fabricated on a 777-nm-thick Si$_3$N$_4$-on-silica platform; the fabrication process is described in the Methods. Figure 3(a) shows an optical image of the chip, a microscope image of the FP microresonator, and a scanning electron micrograph of the chirped Bragg grating. The normalized reflection spectrum (Fig.3(b)) exhibits a series of high-$Q$ resonances within the reflection band. By fitting the resonance line shapes, we extract the intrinsic dissipation and external coupling rates (Fig.3(c)), yielding an intrinsic quality factor of $\approx 4\times10^6$ near the center of the bandgap (around 1550~nm). The external coupling $Q$ is likewise maximized near the band center, ensuring efficient loading of the microresonator. Over the wavelength range from 1530~nm to 1570~nm, the resonances remain within the bandgap and maintain high $Q$.

Dispersion measurements confirm that the FP microresonator provides the targeted anomalous dispersion, with $D_2/2\pi = 90$~MHz. For comparison, a ring microresonator with the same waveguide cross-section (777~nm~$\times$~2~$\mu$m) exhibits a much smaller dispersion of $D_2/2\pi = 0.877$~MHz (Fig.3(d)). The FP architecture therefore enhances the anomalous dispersion by more than two orders of magnitude relative to a conventional ring geometry.

\medskip 
\section{ Comb generation and characterization}

\begin{figure*}[t]
\centering\includegraphics[width=\linewidth]{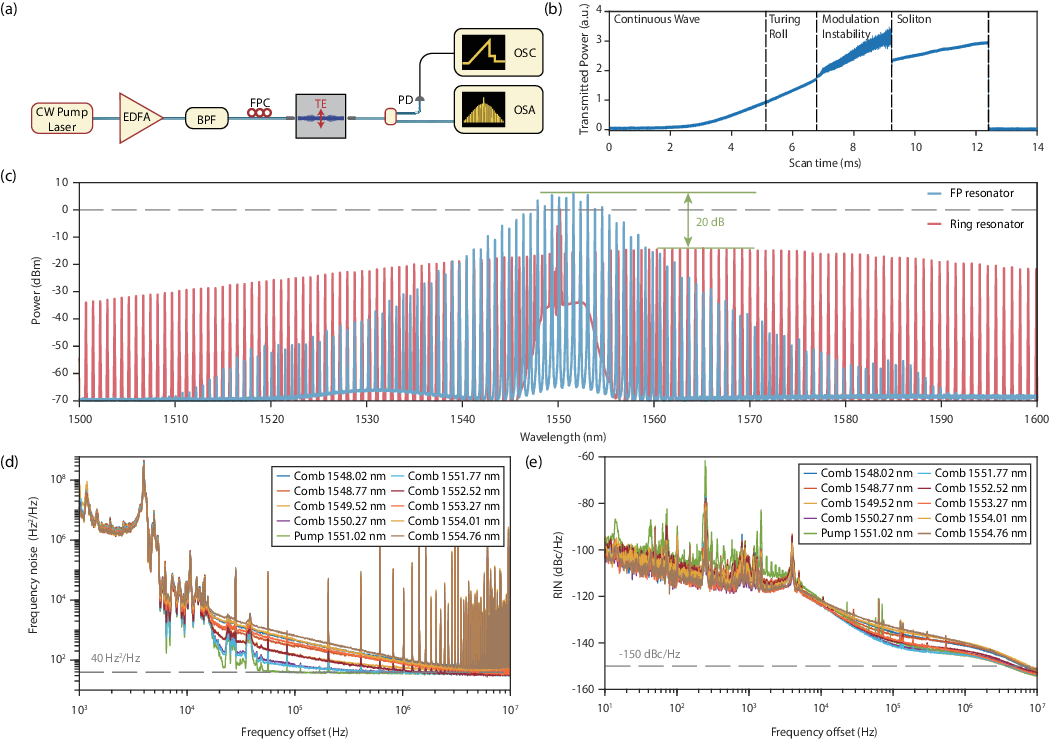}
\caption{{\bf Generation and characterization of the soliton microcombs.} 
(a) Experimental setup for soliton generation and characterization. EDFA, erbium-doped fiber amplifier; BPF, bandpass filter; FPC, fiber polarization controller; PD, photodetector; OSC, oscilloscope; OSA, optical spectral analyzer. (b) Measured transmitted power versus scan time during laser-frequency tuning. Different intracavity optical states are indicated. (c) Optical spectra of solitons generated in the FP microresonator and in a ring microresonator with a similar $Q$ factor and cross-section. (d) Measured single-sideband frequency-noise spectra of selected comb lines. (e) Measured relative intensity noise (RIN) spectra of the same comb lines.}
\end{figure*}

Soliton microcombs are generated using the setup shown in Fig.4(a). A tunable CW laser near 1550~nm is amplified and coupled, in transverse-electric (TE) polarization, into the FP microresonator, providing an on-chip pump power of approximately 410~mW. The transmitted signal is split into two paths: one path is directed to a high-speed photodetector connected to an oscilloscope to monitor the intracavity dynamics during laser-frequency scans, and the other is sent to an optical spectral analyzer to record the comb spectrum (Fig.4(c)). No additional bandpass filter is required for pump suppression because the integrated gratings inherently filter the transmitted pump.

When the pump is tuned into resonance from the blue-detuned side, we observe the characteristic sequence of dynamical states: continuous wave, Turing rolls, modulation instability, and ultimately a low-noise soliton state (Fig.4(b)). The large total energy of the FP-microresonator soliton leads to only a small power change between the modulation instability and soliton states, so the soliton state can be accessed in a thermally stable manner without auxiliary triggering schemes~\cite{yi2015soliton,brasch2016photonic,Zhou2019SolitonBursts}. In transmission, soliton formation manifests as a stable step that appears after the MI regime and persists over a broad detuning range. The soliton state is reached simply by slowly tuning the laser frequency into this step, and we operate near the end of the step to obtain the broadest comb.

The output soliton microcomb exhibits a total on-chip power of 29.2~mW after correcting for the insertion loss between the chip and the output lensed fiber, corresponding to a conversion efficiency of 7.1\%. The ten strongest comb lines each exceed 1~mW, and the strongest comb line reaches $\sim$XX~mW (Fig.4(c)). For comparison, we generate Kerr solitons in a ring microresonator with a similar $Q$ factor and an identical waveguide cross-section. Under comparable pump power and coupling conditions, the power in the central comb lines of the FP microresonator is nearly 20~dB higher than in the ring microresonator, in agreement with the difference in dispersion.

We characterize the coherence of the comb lines in terms of relative intensity noise (RIN) and frequency noise. The RIN is measured by filtering individual comb lines, detecting them with a low-noise photodiode, and analyzing the baseband noise using a commercial phase-noise analyzer. To increase measurement sensitivity, the comb lines are amplified to $\sim$XX~mW before detection. All measured lines exhibit similar RIN spectra, reaching levels below $-150$~dBc/Hz at high offset frequencies (Fig.4(d)). The frequency noise is obtained using a delayed self-heterodyne interferometer, in which the comb line under test is split into a delayed and a frequency-shifted arm and then recombined to form a beat note \cite{jin2021hertz,lao2023quantum}. At high offset frequencies, all comb lines show frequency noise of about $40~\mathrm{Hz^2/Hz}$, corresponding to Lorentzian linewidths of $\sim 250$~Hz, in good agreement with the pump-laser linewidth (Fig.4(e)). At intermediate offset frequencies (10~kHz--1~MHz), comb lines farther from the pump exhibit increased noise, which results from the cascaded contributions of the repetition rate noise.

\begin{figure*}[t]
\centering\includegraphics[width=\linewidth]{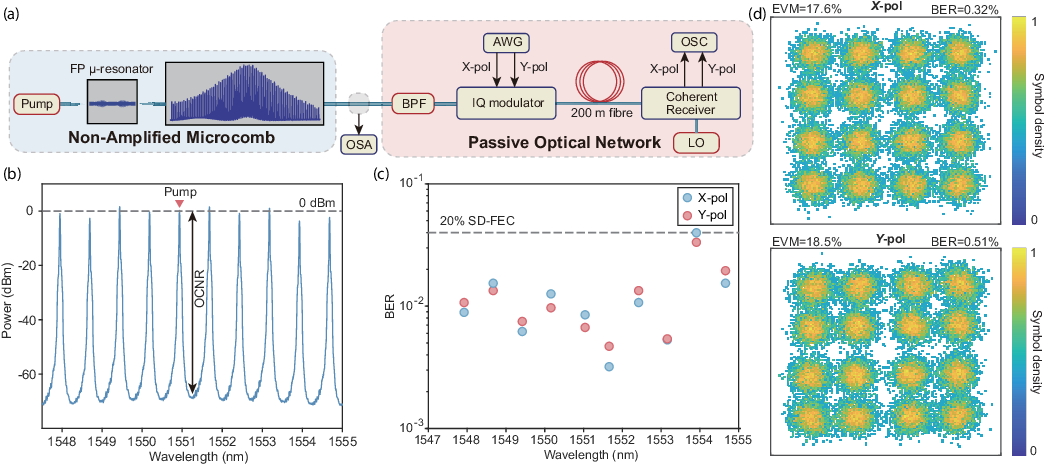}
\caption{{\bf Comb-driven transmission experiments.} 
(a) Experimental setup for passive optical network transmission. AWG, arbitrary waveform generator; LO, local oscillator provided by a tunable laser. (b) Optical spectrum of the ten comb lines used in the transmission experiments, sampled before the bandpass filter. (c) Measured bit-error ratios (BERs) of the data channels in the X polarization (blue) and Y polarization (red). The dashed line denotes a forward-error-correction (FEC) threshold at $4\times10^{-2}$, corresponding to a 20\% soft-decision FEC based on spatially coupled low-density parity-check codes \cite{schuh2017single}. (d) Representative constellation diagrams for both polarizations at a carrier wavelength of 1551.77~nm, with color indicating the relative symbol density.}
\end{figure*}

\medskip 
\section{ Comb-driven passive optical network}

To assess the suitability of the high-power microcomb for optical communication, we perform a transmission experiment based on a passive optical network architecture (Fig.5(a)). The generated microcomb is sent directly to a coherent optical transceiver without optical amplification. The transceivers are realized using silicon-photonic circuits that integrate dual-polarization IQ modulators and balanced photodetectors. Owing to the limited number of available transceivers, we employ a channel-by-channel strategy: a fiber-Bragg-grating bandpass filter selects one comb line at a time as the transmission carrier. Even after chip-to-fiber coupling losses and transmission losses en route to the transmitter (4~dB in total), four comb lines still provide powers above 0~dBm, and all ten carriers exhibit optical carrier-to-noise ratios (OCNRs) exceeding 60~dB when measured with a 40-pm resolution bandwidth (Fig.5(b)). The polarization state is adjusted before modulation to ensure an equal power split between the two polarization axes.

Each selected comb line is encoded with a 16-QAM format at a symbol rate of 25~GBd on both polarizations. The modulated signal is launched into a 200-m single-mode fiber span and then received by a second coherent transceiver. The detected signals are digitized using a real-time oscilloscope and processed offline using a conventional digital signal processing (DSP) flow comprising matched filtering, frequency synchronization, and decision-feedback equalization (Section 4 in Supplement).

The measured bit-error ratios (BERs) for all ten channels and both polarizations are summarized in Fig.5(c), and representative constellation diagrams for the highest-power carrier at 1551.77~nm are shown in Fig.5(d). No optical amplifier is used anywhere in the link, so the BERs correlate directly with the input comb-line powers. The losses induced by the bandpass filter, IQ modulator, and transmission fiber are 1~dB, 9~dB, and 0.5~dB, respectively. The highest input comb-line power is 1.5~dBm at 1551.77~nm, yielding a BER of 0.32\% in the X polarization. The lowest input comb-line power is $-3.7$~dBm at 1554.01~nm, for which a BER of 3.98\% is recorded in the X polarization. All channels achieve BERs below the 20\% soft-decision forward-error-correction (SD-FEC) threshold of $4\times10^{-2}$ reported in Ref.~\cite{schuh2017single}, corresponding to an aggregate data rate of 2~Tb/s.

\medskip 
\section{Discussion}

In summary, we demonstrate that FP microresonators provide a highly versatile platform for dispersion engineering and thereby enable high-power soliton microcombs. Future advances should focus on improving power efficiency~\cite{yang2024efficient} and photonic integration. Coupled-resonator geometries on the FP platform offer a direct means to enhance efficiency, either by introducing a controllable frequency shift~\cite{helgason2023surpassing,wang2025integratedSolitonTurnkey} or by providing resonant enhancement of the pump power~\cite{xue2019super,zhu2025ultra}. For integration with on-chip lasers, directly coupling a DFB laser to the FP microresonator is not viable because of strong reflections from the Bragg gratings. By contrast, using a waveguide coupler \cite{Ulanov2024SILFP} or a microring coupler \cite{zhu2025ultra} can enable practical integration of the FP microresonator with DFB lasers. In this configuration, the intrinsic bidirectional mode structure of the FP microresonator can be harnessed to generate a strong feedback signal to the laser, inducing self-injection locking that narrows the laser linewidth and enables stable, long-term operation without external intervention~\cite{liang2015high,stern2018battery,pavlov2018narrow,raja2019electrically,shen2020integrated,jin2021hertz,wang2025compact,wang2025integratedSolitonTurnkey}.

Despite the coherent link demonstrated in this work, deploying soliton microcombs in practical coherent CPO architectures must address the increased complexity of DSP. This burden can be mitigated by demodulation using coherence-cloned comb sources~\cite{Geng2022Coherent,zhang2023clone}, or by transmitting unmodulated comb signals through multicore fibers and performing modulation and detection in a distributed manner. Such approaches support higher aggregate data rates without incurring prohibitive DSP overhead. While the present experiments already demonstrate 2~Tb/s transmission over a single fiber, scaling beyond 5~Tb/s should be feasible if the comb bandwidth is further broadened using more efficient pumping schemes~\cite{yang2024efficient}. We thereby anticipate the use of FP-based high-power soliton microcombs as enabling sources for next-generation CPO systems operating at petabit-per-second scales~\cite{rizzo2022petabit,jorgensen2022petabit}.

\bibliography{ref}

\medskip

\medskip

\noindent\textbf{Acknowledgments}

\begin{footnotesize}
\noindent This work was supported by Quantum Science and Technology-National Science and Technology Major Project (Grants No. 2021ZD0301500). The authors thank Jincheng Li, Zhigang Hu, Hao Yang, Ruokai Zheng, and Xiaoxuan Peng for assistance in fabrication. The fabrication in this work was supported by the Peking University Nano-Optoelectronic Fabrication Center, Micro/nano Fabrication Laboratory of Synergetic Extreme Condition User Facility (SECUF), Songshan Lake Materials Laboratory, and the Advanced Photonics Integrated Center of Peking University. 
\end{footnotesize}
\medskip

\noindent\textbf{Disclosure}

\begin{footnotesize}
\noindent The authors declare no competing interests.
\end{footnotesize}

\medskip
\noindent\textbf{Data availability}
\begin{footnotesize}
\noindent The data that support the plot within this paper and other findings of this study are available upon publication. 
\end{footnotesize}

%%%%%%%%%% If using BibTeX:

%%%%%%%%%% If preparing manually:
% \begin{thebibliography}{1}
% \newcommand{\enquote}[1]{``#1''}

% \bibitem{Zhang:14}
% Y.~Zhang, S.~Qiao, L.~Sun, Q.~W. Shi, W.~Huang, L.~Li, and Z.~Yang,
%   \enquote{Photoinduced active terahertz metamaterials with nanostructured
%   vanadium dioxide film deposited by sol-gel method,}
%   {\protect\JournalTitle{Optics Express}} \textbf{22}, 11070--11078 (2014).

% \bibitem{Optica}
% {Optica}, \enquote{{Optica Publishing Group},}
%   \url{http://www.opg.optica.org}.

% \bibitem{FORSTER2007}
% P.~Forster, V.~Ramaswamy, P.~Artaxo, T.~Bernsten, R.~Betts, D.~Fahey,
%   J.~Haywood, J.~Lean, D.~Lowe, G.~Myhre, J.~Nganga, R.~Prinn, G.~Raga,
%   M.~Schulz, and R.~V. Dorland, \enquote{Changes in atmospheric consituents and
%   in radiative forcing,} in \enquote{Climate Change 2007: The Physical Science
%   Basis. Contribution of Working Group 1 to the Fourth Assesment Report of
%   Intergovernmental Panel on Climate Change,}  S.~Solomon, D.~Qin, M.~Manning,
%   Z.~Chen, M.~Marquis, K.~B. Averyt, M.~Tignor, and H.~L. Miler, eds.
%   (Cambridge University Press, 2007).

% \end{thebibliography}

\end{document}